\documentclass[aps,pra,twocolumn,reprint,amsmath,amssymb,superscriptaddress,floatfix,footinbib,longbibliography]{revtex4-1}

\usepackage{graphicx}
\usepackage{epstopdf}

\usepackage[T1]{fontenc}
\usepackage[applemac]{inputenc}
\usepackage{lmodern}
\usepackage[english]{babel}

\usepackage{ae}
\usepackage{units}

\usepackage{amsmath,amssymb,natbib,bm}
\usepackage{psfrag}
\usepackage{subfigure}
\usepackage{amsthm}

\usepackage{fixmath}
\usepackage{booktabs}

\usepackage{slashed}

\usepackage[americaninductors]{circuitikz}
\usepackage{tikz}
\usetikzlibrary{arrows}

\usepackage{color}
\usepackage{url}

\usepackage[colorlinks]{hyperref}
\hypersetup{%
        plainpages=true,
        breaklinks=true,
        hypertexnames=false,
        pageanchor=true,
        colorlinks=true,
        linkcolor={blue},
        citecolor={magenta},
        urlcolor={blue},
        anchorcolor={black}
      }

\newcommand{\figref}[1]{\mbox{Fig.~\ref{#1}}}

\newcommand{\secref}[1]{\mbox{Sec.~\ref{#1}}}

\newcommand{\appref}[1]{\mbox{Appendix~\ref{#1}}}
\renewcommand{\eqref}[1]{\mbox{Eq.~(\ref{#1})}}

\newcommand{\ket}[1]{|#1\rangle}

\newcommand{\ketbra}[2]{\left| #1 \rangle \langle #2 \right|}
\newcommand{\brakket}[3]{\left\langle #1\left| #2 \right| #3\right\rangle}
\newcommand{\expec}[1]{\left\langle #1 \right\rangle}

\newcommand{\comm}[2]{\left[ #1, #2 \right]}
\newcommand{\lind}[1]{\mathcal{D}\left[#1\right]}

\newcommand{\sz}{\hat \sigma_z}
\newcommand{\sx}{\hat \sigma_x}
\newcommand{\sm}{\hat \sigma_-}
\renewcommand{\sp}{\hat \sigma_+}

\newcommand{\abs}[1]{\left|#1\right|}

\newcommand{\abssq}[1]{\left| #1 \right|^2}

\newcommand{\nn}{\nonumber}

\newcommand{\be}{\begin{equation}}
\newcommand{\ee}{\end{equation}}
\newcommand{\bea}{\begin{eqnarray}}
\newcommand{\eea}{\end{eqnarray}}

\begin{document}

\title{Frequency conversion in ultrastrong cavity QED}

\author{Anton Frisk Kockum}
\email[e-mail:]{anton.frisk.kockum@gmail.com}
\affiliation{Center for Emergent Matter Science, RIKEN, Saitama 351-0198, Japan}

\author{Vincenzo Macrì}
\affiliation{Dipartimento di Fisica e di Scienze della Terra, Universit\`{a} di Messina, I-98166 Messina, Italy}
\affiliation{Center for Emergent Matter Science, RIKEN, Saitama 351-0198, Japan}

\author{Luigi Garziano}
\affiliation{School of Physics and Astronomy, University of Southampton, Southampton, SO17 1BJ, United Kingdom}
\affiliation{Dipartimento di Fisica e di Scienze della Terra, Universit\`{a} di Messina, I-98166 Messina, Italy}
\affiliation{Center for Emergent Matter Science, RIKEN, Saitama 351-0198, Japan}

\author{Salvatore Savasta}
\affiliation{Dipartimento di Fisica e di Scienze della Terra, Universit\`{a} di Messina, I-98166 Messina, Italy}
\affiliation{Center for Emergent Matter Science, RIKEN, Saitama 351-0198, Japan}

\author{Franco Nori}
\affiliation{Center for Emergent Matter Science, RIKEN, Saitama 351-0198, Japan}
\affiliation{Physics Department, The University of Michigan, Ann Arbor, Michigan 48109-1040, USA}

\date{\today}

\begin{abstract}

We propose a new method for frequency conversion of photons which is both versatile and deterministic. We show that a system with two resonators ultrastrongly coupled to a single qubit can be used to realize both single- and multiphoton frequency-conversion processes. The conversion can be exquisitely controlled by tuning the qubit frequency to bring the desired frequency-conversion transitions on or off resonance. Considering recent experimental advances in ultrastrong coupling for circuit QED and other systems, we believe that our scheme can be implemented using available technology.

\end{abstract}


\maketitle

\section{Introduction}

Frequency conversion in quantum systems~\cite{Kumar1990, Huang1992} is important for many quantum technologies. The optimal working points of devices for transmission, detection, storage, and processing of quantum states are spread across a wide spectrum of frequencies~\cite{OBrien2009, Buluta2011}. Interfacing the best of these devices is necessary to create quantum networks~\cite{Kimble2008} and other powerful combinations of quantum hardware. Examples of frequency-conversion setups developed for such purposes include upconversion for photon detection~\cite{Albota2004} and storage~\cite{Tanzilli2005}, since both these things are easier to achieve at a higher frequency than what is optimal for telecommunications. Downconversion in this frequency range has also been demonstrated~\cite{Ou2008, Ding2010, Takesue2010}, and recently even strong coupling between a telecom and a visible optical mode~\cite{Guo2016}. Additionally, advances in quantum information processing with superconducting circuits at microwave frequencies~\cite{You2011, Devoret2013} is driving progress on frequency conversion between optical and microwave frequencies~\cite{Bochmann2013, Andrews2014, Shumeiko2016, Hisatomi2016}.

Circuit quantum electrodynamics (QED)~\cite{You2003, Wallraff2004, Blais2004, You2011, Xiang2013} offers a wealth of possibilities for frequency conversion at microwave frequencies; some of these schemes can also be generalized to optical frequencies. By modulating the magnetic flux through a superconducting quantum interference device (SQUID) in a transmission line resonator, the frequency of the photons in the resonator can be changed rapidly~\cite{Wallquist2006, Sandberg2008, Johansson2010} or two modes of the resonator can be coupled~\cite{Chirolli2010, Zakka-Bajjani2011}. Other driven Josephson-junction-based devices can also be used for microwave frequency conversion~\cite{Abdo2013b, Kamal2014}. Downconversion has been proposed for setups with $\Delta$-type three-level atoms~\cite{Marquardt2007, Koshino2009, Sanchez-Burillo2016} and demonstrated with an effective three-level $\Lambda$ system~\cite{Inomata2014}. Upconversion of a two-photon drive has been shown for a flux qubit coupled to a resonator in a way that breaks parity symmetry~\cite{Deppe2008}. Indeed, the $\Delta$-type level structure in a flux qutrit \cite{Liu2005} even makes possible general three-wave mixing \cite{Liu2014}. Recently, frequency conversion was also demonstrated for two sideband-driven microwave $LC$-resonators coupled through a mechanical resonator~\cite{Lecocq2016}.

The approach to frequency conversion that we propose in this article is based on two cavities or resonator modes coupled ultrastrongly to a two-level atom (qubit). The regime of ultrastrong coupling (USC), where the coupling strength starts to become comparable to the bare transition frequencies in the system, has only recently been reached in a number of solid-state systems~\cite{Gunter2009, Forn-Diaz2010, Niemczyk2010, Todorov2010, Schwartz2011, Scalari2012, Geiser2012, Kena-Cohen2013, Gambino2014, Maissen2014, Goryachev2014, Baust2016, Forn-Diaz2017, Yoshihara2017, Chen2016, George2016, Langford2016, Braumuller2016, Yoshihara2016}. Among these, a few circuit-QED experiments provide some of the clearest examples~\cite{Forn-Diaz2010, Niemczyk2010, Baust2016, Forn-Diaz2017, Yoshihara2017, Chen2016, Langford2016, Braumuller2016, Yoshihara2016}, including the largest coupling strength reported~\cite{Yoshihara2017}. While the USC regime displays many striking physical phenomena~\cite{DeLiberato2007, Ashhab2010, Cao2010, Cao2011, Stassi2013, Sanchez-Burillo2014, DeLiberato2014, Lolli2015, DiStefano2016, Cirio2016}, we are here only concerned with the fact that it enables higher-order processes that do not conserve the number of excitations in the system, an effect which has also been noted for a multilevel atom coupled to a resonator~\cite{Zhu2013}. Examples of such processes include multiphoton Rabi oscillations~\cite{Ma2015, Garziano2015} and a single photon exciting multiple atoms~\cite{Garziano2016}. Indeed, almost any analogue of processes from nonlinear optics is feasible~\cite{Kockum2017a}; this can be regarded as an example of quantum simulation~\cite{Buluta2009, Georgescu2014}. Just like the analytical solution for the quantum Rabi model~\cite{Braak2011} is now being extended to multiple qubits~\cite{Braak2013, Peng2013} and multiple resonators~\cite{Chilingaryan2015, Duan2015, Alderete2016}, we here extend the exploration of non-excitation-conserving processes to multiple resonators.

In our proposal, the qubit frequency is tuned to make various frequency-converting transitions resonant. For example, making the energy of a single photon in the first resonator equal to the sum of the qubit energy and the energy of a photon in the second resonator enables the conversion of the former (a high-energy photon) into the latter (a low-energy photon plus a qubit excitation) and vice versa. In the same way, a single photon in the first resonator can be converted into multiple photons in the second resonator (and vice versa) if the qubit energy is tuned to make such a transition resonant. The proposed frequency-conversion scheme is deterministic and allows for a variety of different frequency-conversion processes in the same setup. The setup should be possible to implement in state-of-the-art circuit QED, but the idea also applies to other cavity QED systems. 

We note that the process of parametric down-conversion in this type of circuit-QED setup has been considered previously~\cite{Moon2005}, but in a regime of weaker coupling and without using the qubit to control the process. Also, it has been shown that a beamsplitter-type coupling between two resonators can be controlled by changing the qubit state~\cite{Mariantoni2008} or induced for weaker qubit-resonator coupling by driving the qubit~\cite{Prado2006}, but the proposal presented here offers greater versatility and simplicity for frequency conversion.

This article is organized as follows. In \secref{sec:Model}, we define the system under consideration and explain the principle behind frequency conversion based on USC. In Secs.~\ref{sec:SinglePhoton} and \ref{sec:MultiPhoton}, we show the details of the single- and multiphoton frequency-conversion processes, respectively, including both analytical and numerical calculations. We conclude and give an outlook for future work in \secref{sec:SummaryOutlook}. Details of some analytical calculations are given in \appref{app:AnalyticalCalculations}.

\section{Model}
\label{sec:Model}

\begin{figure}
\centering
\includegraphics[width=\columnwidth]{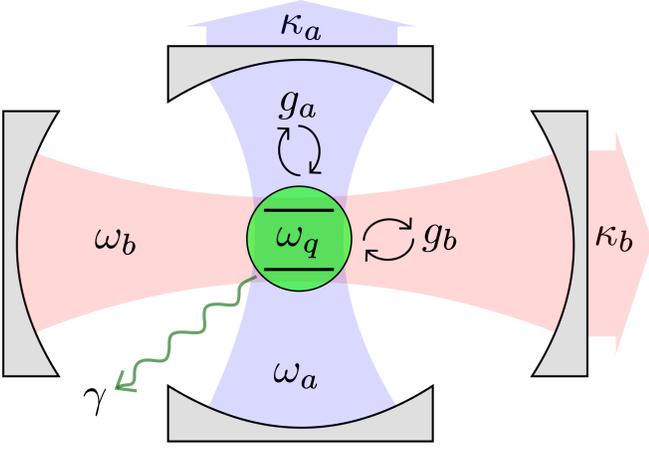}
\caption{A sketch of the system. A qubit (green) is coupled to two resonator modes (blue, $a$, and red, $b$). Decoherence channels for the qubit (relaxation rate $\gamma$) and the resonators (relaxation rates $\kappa_a$, $\kappa_b$) are included. \label{fig:Setup}}
\end{figure}

We consider a setup where a qubit with transition frequency $\omega_q$ is coupled to two resonators with resonance frequencies $\omega_a$ and $\omega_b$, respectively, as sketched in \figref{fig:Setup}. The Hamiltonian is ($\hbar = 1$)
\bea
\hat H &=& \omega_a \hat a^\dag \hat a + \omega_b \hat b^\dag \hat b + \frac{\omega_q}{2}\sz \nn\\
&&+ \left[g_a \left(\hat a + \hat a^\dag \right) + g_b \left(\hat b + \hat b^\dag \right) \right]\nn\\
&&\times \left(\sx \cos\theta + \sz\sin\theta \right),
\label{eq:Ham}
\eea
where $g_a$ ($g_b$) denotes the strength of the coupling between the qubit and the first (second) resonator. The creation and annihilation operators for photons in the first (second) resonator are $\hat a$ and $\hat a^\dag$ ($\hat b$ and $\hat b^\dag$), respectively. The angle $\theta$ parameterizes the amount of longitudinal and transverse coupling as, for example, in experiments with flux qubits~\cite{You2007, Deppe2008, Niemczyk2010, Forn-Diaz2010, Baust2016, Yoshihara2016}; $\sx$ and $\sz$ are Pauli matrices for the qubit. 

Note that we do not include a direct coupling between the two resonators. Such a coupling is seen in experiments~\cite{Baust2016}, but here we will only be concerned with situations where the resonators are far detuned from each other, meaning that this coupling term can safely be neglected. Likewise, we do not include higher modes of the resonators. While they may contribute in experiments with cavities and transmission-line resonators, they can be avoided by using lumped-element resonators~\cite{Kim2011, Yoshihara2016}.

The crucial feature of \eqref{eq:Ham} for our frequency-conversion scheme is that some of the coupling terms do not conserve the number of excitations in the system. The $\sz$ coupling terms act to change the photon number in one of the resonators by one, while keeping the number of qubit excitations unchanged. Likewise, the $\sx$ coupling contains terms like $\hat a\sm$ and $\hat b^\dag \sp$ that change the number of excitations in the system by two. For weak coupling strengths, all such terms can be neglected using the rotating-wave approximation (RWA), but in the USC regime the higher-order processes that these terms enable can become important and function as second- or third-order nonlinearities in nonlinear optics~\cite{Kockum2017a}. 

To include the effect of decoherence in our system, we use a master equation on the Lindblad form in our numerical simulations. The master equation reads
\be
\dot{\hat \rho} = - i \comm{\hat H}{\hat \rho} + \sum_{j,k > j} \left( \Gamma_a^{jk} + \Gamma_b^{jk} + \Gamma_q^{jk} \right)\lind{\ketbra{j}{k}} \hat \rho,
\ee
where $\hat \rho$ is the density matrix of the system, $\lind{\hat c}\rho = \hat c \hat \rho \hat c^\dag - \frac{1}{2} \hat \rho \hat c^\dag \hat c - \frac{1}{2} \hat c^\dag \hat c \hat \rho$, and the states in the sum are eigenstates of the USC system. The relaxation rates are given by $\Gamma_a^{jk} = \kappa_a \abssq{X_a^{jk}}$, $\Gamma_b^{jk} = \kappa_b \abssq{X_b^{jk}}$, and $\Gamma_q^{jk} = \gamma \abssq{C^{jk}}$, where $c^{jk} = \brakket{j}{\hat c}{k}$ with $\hat X_a = \hat a + \hat a^\dag$, $\hat X_b = \hat b + \hat b^\dag$, and $\hat C = \sx$~\cite{Beaudoin2011, Ridolfo2012}. Writing the master equation in the eigenbasis of the full system avoids unphysical effects, such as emission of photons from the ground state. Similarly, to correctly count the number of photonic and qubit excitations we use $\expec{\hat X_a^- \hat X_a^+}$, $\expec{\hat X_b^- \hat X_b^+}$, and $\expec{\hat C^- \hat C^+}$, where the plus and minus signs denote the positive and negative frequency parts, respectively, of the operators in the system eigenbasis, instead of $\expec{\hat a^\dag \hat a}$, $\expec{\hat b^\dag \hat b}$, and $\expec{\sp \sm}$~\cite{Ridolfo2012}.

\section{Single-photon frequency conversion}
\label{sec:SinglePhoton}

We first consider single-photon frequency conversion, where one photon in the first resonator is converted into one photon of a different frequency in the second resonator, or vice versa. The conversion is aided by the qubit. Without loss of generality, we take $\omega_a > \omega_b$. For the conversion to work, we then need $\omega_a \approx \omega_b + \omega_q$, such that the states $\ket{1,0,g}$ and $\ket{0,1,e}$ are close to resonant. Due to the presence of longitudinal coupling in the Hamiltonian in \eqref{eq:Ham}, transitions between these two states are possible even though their excitation numbers and parity differ. 

\begin{figure}
\centerline{
  \resizebox{\columnwidth}{!}{
    \begin{tikzpicture}[
      scale=0.75,
      level/.style={thick},
      virtual/.style={thick,densely dashed},
      ztrans/.style={thick,->,shorten >=0.2cm,shorten <=0.2cm,>=stealth,densely dashed,color=red},
      nrtrans/.style={thick,->,shorten >=0.2cm,shorten <=0.2cm,>=stealth,densely dashed,color=blue},
      rtrans/.style={thick,->,shorten >=0.2cm,shorten <=0.2cm,>=stealth,color=blue},
      classical/.style={thin,double,->,shorten >=4pt,shorten <=4pt,>=stealth}
    ]
    \draw[level] (2cm,0cm) -- (0cm,0cm) node[midway,below] {$\ket{0,0,g}$};
    \draw[level] (2cm,1cm) -- (0cm,1cm) node[midway,below] {$\ket{0,0,e}$};
    \draw[level] (-0.5cm,3cm) -- (-2.5cm,3cm) node[left] {$\ket{1,0,g}$};
    \draw[level] (2.5cm,3cm) -- (4.5cm,3cm) node[right] {$\ket{0,1,e}$};
    \draw[level] (2cm,5cm) -- (0cm,5cm) node[midway,above] {$\ket{1,1,g}$};
    \draw[level] (2cm,6cm) -- (0cm,6cm) node[midway,above] {$\ket{1,1,e}$};
    \draw[nrtrans] (-2cm,3cm) -- (0.4cm,6cm) node[midway, left] {$g_b\cos\theta$};
    \draw[ztrans] (-1.5cm,3cm) -- (0.9cm,5cm) node[midway, right, xshift = -0.35cm, yshift = -0.3cm] {$-g_b\sin\theta$};
    \draw[rtrans] (-1.5cm,3cm) -- (0.9cm,1cm) node[midway, right, xshift = -0.3cm, yshift = 0.35cm] {$g_a\cos\theta$};    
    \draw[ztrans] (-2cm,3cm) -- (0.4cm,0cm) node[midway, left, xshift = -0.1cm] {$-g_a\sin\theta$};
    \draw[ztrans] (1.6cm,6cm) -- (4cm,3cm) node[midway, right, xshift = 0.1cm] {$g_a\sin\theta$};
    \draw[rtrans] (1.1cm,5cm) -- (3.5cm,3cm) node[midway, left, xshift = 0.25cm, yshift = -0.3cm] {$g_a\cos\theta$};
    \draw[ztrans] (1.1cm,1cm) -- (3.5cm,3cm) node[midway, left, xshift = 0.3cm, yshift = 0.35cm] {$g_b\sin\theta$};
    \draw[nrtrans] (1.6cm,0cm) -- (4cm,3cm) node[midway, right] {$g_b\cos\theta$};
    \end{tikzpicture}
  }
}
\caption{The four lowest-order processes contributing to a transition between $\ket{1,0,g}$ and $\ket{0,1,e}$. For this illustration, the parameter values $\omega_a = 3\omega_q$ and $\omega_b = 2\omega_q$ were used to set the positions of the energy levels. The transitions that do not conserve excitation number are shown as dashed lines, and the excitation-number-conserving transitions are shown as solid lines. Red lines correspond to $\sz$ (longitudinal) coupling and blue lines to $\sx$ (transverse) coupling in the Hamiltonian given in \eqref{eq:Ham}. Each transition is labelled by its matrix element. \label{fig:Processes10gTo01e}}
\end{figure}
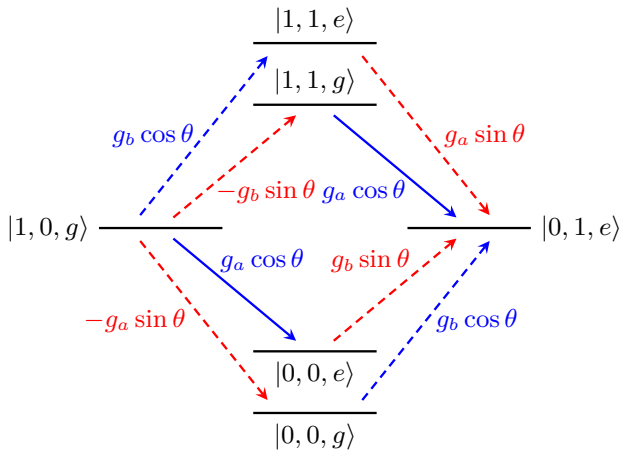

The intermediate states and transitions contributing (in lowest order) to the $\ket{1,0,g} \leftrightarrow \ket{0,1,e}$ transition are shown in \figref{fig:Processes10gTo01e}. Virtual transitions to and from one of the four intermediate states $\ket{0,0,g}$, $\ket{0,0,e}$, $\ket{1,1,g}$, and $\ket{1,1,e}$ connect $\ket{1,0,g}$ and $\ket{0,1,e}$ in two steps. This is the minimum number of steps possible, since the terms in the Hamiltonian in \eqref{eq:Ham} can only create or annihilate a single photon at a time. From the figure, it is also clear that no path exists between $\ket{1,0,g}$ and $\ket{0,1,e}$ that does not involve longitudinal coupling (dashed red arrows in the figure). 

To calculate the effective coupling between the states $\ket{1,0,g}$ and $\ket{0,1,e}$, we truncate the Hamiltonian from \eqref{eq:Ham} to the six states shown in \figref{fig:Processes10gTo01e}. Written on matrix form, this truncated Hamiltonian becomes
\begin{widetext}
\bea
\hat H = 
\begin{pmatrix}
- \frac{\omega_q}{2} & 0 & - g_a\sin\theta & g_b\cos\theta & 0 & 0 \\
0 & \frac{\omega_q}{2} & g_a\cos\theta & g_b\sin\theta & 0 & 0 \\
- g_a\sin\theta & g_a\cos\theta & \omega_a - \frac{\omega_q}{2} & 0 & - g_b\sin\theta & g_b\cos\theta \\
g_b\cos\theta & g_b\sin\theta & 0 & \omega_b + \frac{\omega_q}{2} & g_a\cos\theta & g_a\sin\theta \\
0 & 0 & - g_b\sin\theta & g_a\cos\theta & \omega_a + \omega_b - \frac{\omega_q}{2} & 0 \\
0 & 0 & g_b\cos\theta & g_a\sin\theta & 0 & \omega_a + \omega_b + \frac{\omega_q}{2}
\end{pmatrix},
\label{eq:HTruncatedSinglePhoton}
\eea
\end{widetext}
where the states are ordered from left to right as $\ket{0,0,g}$, $\ket{0,0,e}$, $\ket{1,0,g}$, $\ket{0,1,e}$, $\ket{1,1,g}$, and $\ket{1,1,e}$. When the condition $\omega_a \approx \omega_b + \omega_q$ is satisfied, the four intermediate states $\ket{0,0,g}$, $\ket{0,0,e}$, $\ket{1,1,g}$, and $\ket{1,1,e}$ can be adiabatically eliminated. This calculation, shown in \appref{app:AnalyticalCalculationsSinglePhoton}, gives an effective Hamiltonian with a coupling term
\be
\hat H_{\rm c, eff} = g_{\rm eff} \left( \ketbra{1,0,g}{0,1,e} + \ketbra{0,1,e}{1,0,g} \right),
\ee
where the effective coupling between the states $\ket{1,0,g}$ and $\ket{0,1,e}$ has the magnitude
\be
g_{\rm eff} = g_a g_b\sin2\theta \left( \frac{1}{\omega_b} - \frac{1}{\omega_a}\right)
\label{eq:GeffSinglePhoton}
\ee
on resonance. Compared to the direct resonator-qubit coupling in \eqref{eq:Ham}, $g_{\rm eff}$ is weaker by a factor of order $g/\omega$, which is why we need to at least approach the USC regime to observe the single-photon frequency conversion. We note that the effective coupling is maximized when the longitudinal and transverse coupling terms in \eqref{eq:Ham} have equal magnitude. Interestingly, \eqref{eq:GeffSinglePhoton} suggests that frequency conversion can be more efficient if $\omega_b \ll \omega_a$. However, going too far in this direction violates the assumptions behind the adiabatic approximation, which relies on $g_a, g_b \ll \omega_a, \omega_b$.

\begin{figure*}
\centering
\includegraphics[width=\linewidth]{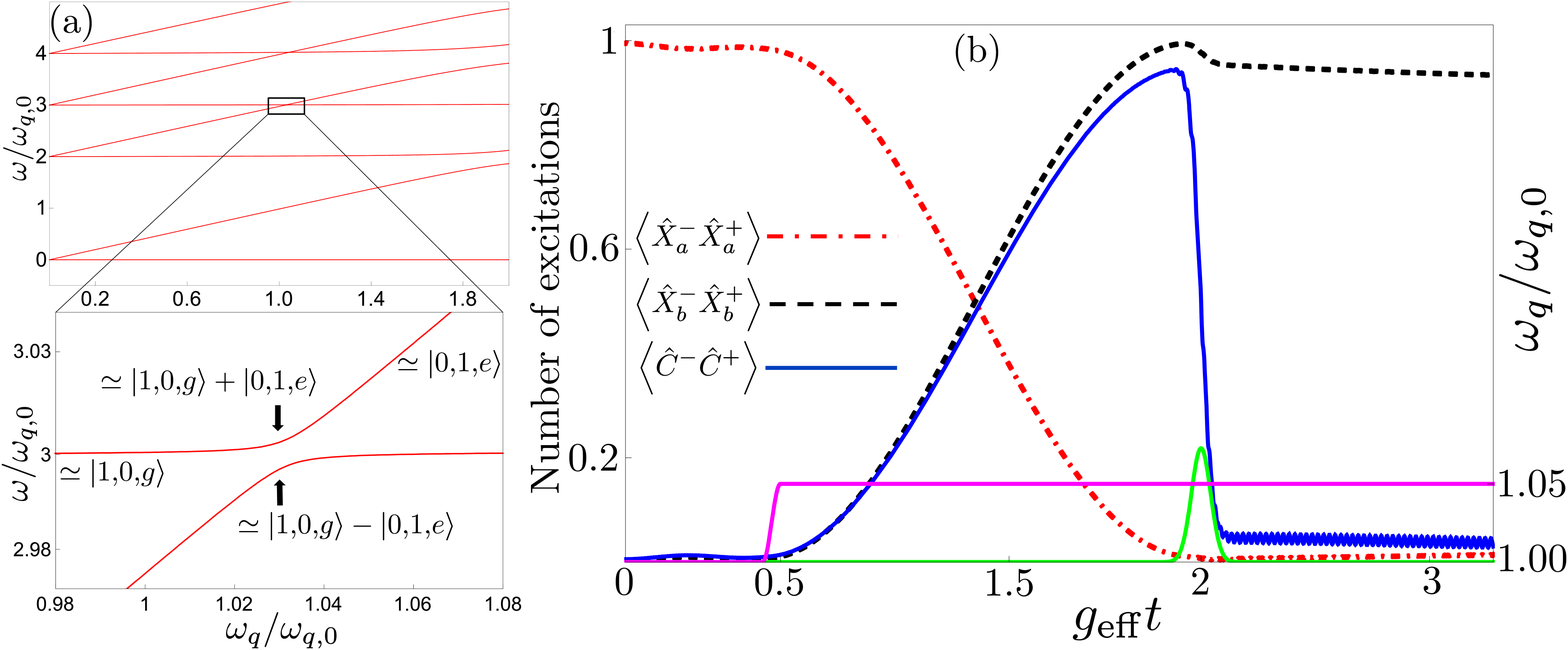}
\caption{Two frequency conversion methods. (a) The figure shows the energy levels of our system plotted as a function of the qubit frequency $\omega_q$, using the parameters $g_a = g_b = 0.15\omega_{q,0}$, $\theta = \pi/6$, $\omega_a = 3\omega_{q,0}$, and $\omega_b = 2\omega_{q,0}$, where $\omega_{q,0}$ is a reference point for the qubit frequency, set such that $\omega_a = \omega_b + \omega_{q,0}$. In the zoom-in, close to the resonance $\omega_a = \omega_b + \omega_q$, we see the anticrossing between $\ket{1,0,g}$ and $\ket{0,1,e}$ with splitting $2g_{\rm eff}$. Up- or down-conversion of single photons can be achieved by adiabatically tuning $\omega_q$ to follow one of the energy levels in the figure from $\ket{1,0,g}$ to $\ket{0,1,e}$, or vice versa. (b) A rapid frequency conversion can be achieved by starting in $\ket{1,0,g}$, far from the resonance $\omega_a = \omega_b + \omega_q$, tuning the qubit frequency (pink solid line) into resonance for half a Rabi period ($\pi/2g_{\rm eff}$) and then sending a pulse (green solid line) to deexcite the qubit. The figure shows the number of excitations in the two resonators (red dashed-dotted line for $a$, black dashed line for $b$) and the qubit (blue solid line) during such a process, including decoherence in the form of relaxation from the resonators and the qubit. The parameters used for the decoherence are $\kappa_a = \kappa_b = \gamma = 4\times 10^{-5} \omega_{q,0}$. 
\label{fig:LevelsAndDynamicTransferSimSinglePhoton}}
\end{figure*}

The existence of this effective coupling suggests at least two ways to perform single-photon frequency conversion. The first is to use adiabatic transfer, starting in $\ket{1,0,g}$ ($\ket{0,1,e}$) with the qubit frequency sufficiently far detuned from the resonance $\omega_a = \omega_b + \omega_q$ and then slowly (adiabatically) changing the qubit frequency until the system ends up in the state $\ket{0,1,e}$ ($\ket{1,0,g}$), following one of the energy levels shown in \figref{fig:LevelsAndDynamicTransferSimSinglePhoton}(a). In this way, a single photon in the first (second) resonator is deterministically down-converted (up-converted) to a single photon of lower (higher) frequency in the second (first) resonator. We note that such adiabatic transfer has been used for robust single-photon generation in circuit QED, tuning the frequency of a transmon qubit to achieve the transition $\ket{0,e} \to \ket{1,g}$~\cite{Johnson2010}. It has also been suggested as a method to generate multiple photons from a single qubit excitation in the USC regime of the standard quantum Rabi model~\cite{Ma2015}.

The second approach, exemplified by a simulation including decoherence in \figref{fig:LevelsAndDynamicTransferSimSinglePhoton}(b), is to initialize the system in one of the states $\ket{1,0,g}$ or $\ket{0,1,e}$, far from the frequency-conversion resonance such that the effective coupling is negligible, quickly tune the qubit into resonance for the duration of half a Rabi oscillation period (set by the effective coupling to be $\pi/2g_{\rm eff}$), and then detune the qubit again (or send a pulse to deexcite it) to turn off the effective interaction. This type of scheme is, for example, commonly used for state transfer between resonators and/or qubits in circuit QED~\cite{Sillanpaa2007, Hofheinz2008, Hofheinz2009, Wang2011, Mariantoni2011}. Letting the resonance last shorter or longer times, any superposition of $\ket{1,0,g}$ or $\ket{0,1,e}$ can be created. The potential for creating superpositions of photons of different frequencies (similar to Ref.~\cite{Zakka-Bajjani2011}) with such a method will be explored in future work.

\begin{figure}
\centering
\includegraphics[width=\columnwidth]{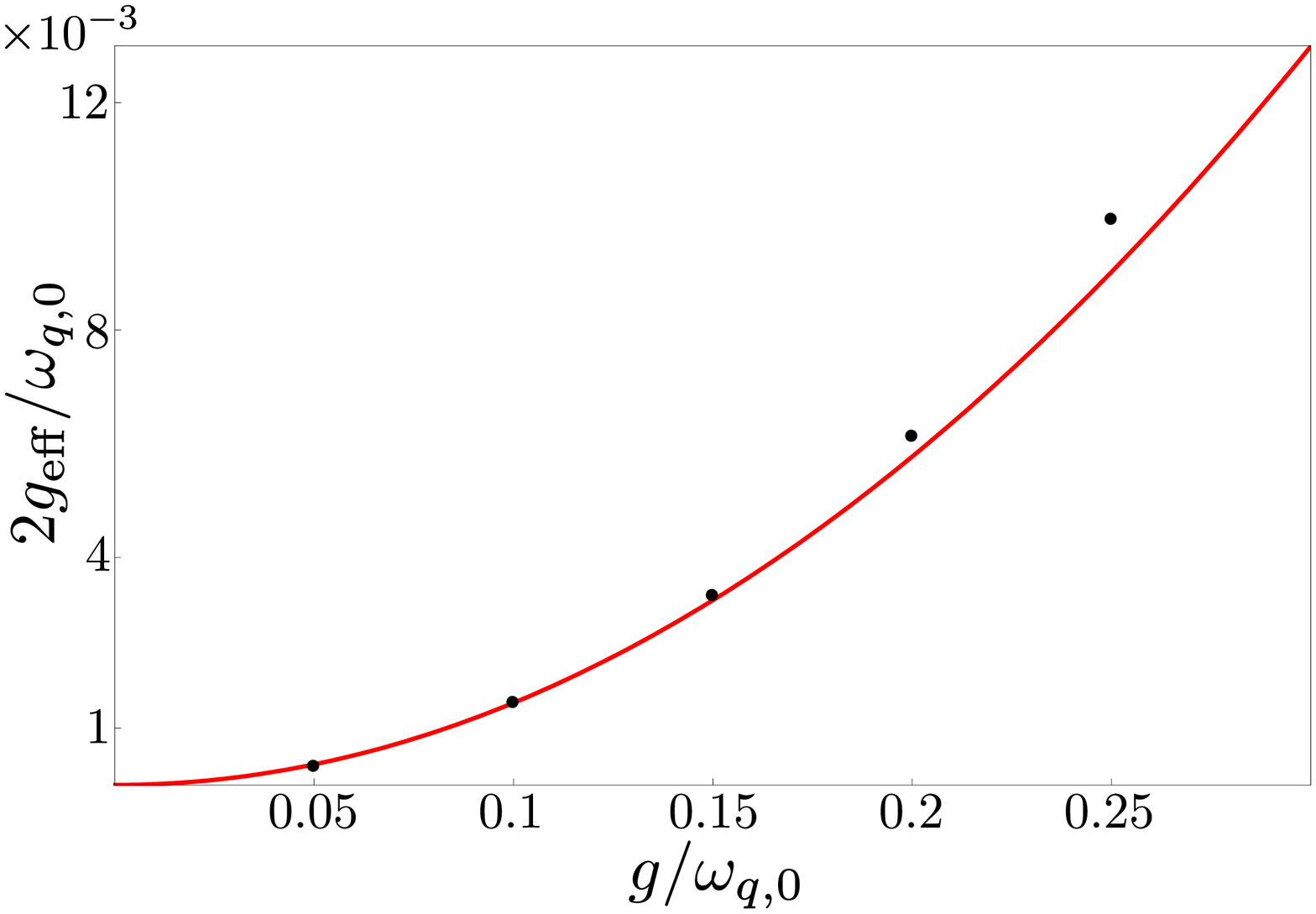}
\caption{Comparison of analytical (red line) and numerical (black dots) results for the effective coupling between the states $\ket{1,0,g}$ and $\ket{0,1,e}$. The graph shows the minimum energy splitting $2g_{\rm eff}/\omega_{q,0}$ as a function of $g/\omega_{q,0}$, where $g = g_a = g_b$, using the same parameters as in \figref{fig:LevelsAndDynamicTransferSimSinglePhoton}. 
\label{fig:ComparingCoupling10g01e}}
\end{figure}

Since the relevant timescales for both these approaches are determined by $g_{\rm eff}$, it is important to know in which parameter range the expression for $g_{\rm eff}$ given in \eqref{eq:GeffSinglePhoton} remains a good approximation. In \figref{fig:ComparingCoupling10g01e}, we show that the expression is valid up to at least $g_a =  g_b = 0.2\omega_{q,0}$ for the parameters used in \figref{fig:LevelsAndDynamicTransferSimSinglePhoton}.

\section{Multi-photon frequency conversion}
\label{sec:MultiPhoton}

We now turn to multi-photon frequency conversion, where, aided by the qubit, one photon in the first resonator is converted into two photons in the second resonator, or vice versa. We continue to adopt the convention that $\omega_a > \omega_b$. In contrast to the single-photon frequency conversion case in \secref{sec:SinglePhoton}, there are now two possibilities for how the qubit state can change during the conversion process. Below, we will study both $\ket{1,0,g} \leftrightarrow \ket{0,2,e}$ and $\ket{1,0,e} \leftrightarrow \ket{0,2,g}$. Since we wish to use the qubit to control the process, we do not consider the process $\ket{1,0,g} \leftrightarrow \ket{0,2,g}$, which to some extent was already included in Ref.~\cite{Moon2005}.

\subsection{$\ket{1,0,g} \leftrightarrow \ket{0,2,e}$}

For the process $\ket{1,0,g} \leftrightarrow \ket{0,2,e}$, we first of all note one more difference compared to the single-photon frequency conversion case in \secref{sec:SinglePhoton}: it changes the number of excitations from 1 to 3, which means that excitation-number parity is conserved. This makes the longitudinal coupling of \eqref{eq:Ham} redundant for achieving the conversion, and to simplify our calculations we therefore hereafter work with the standard quantum Rabi Hamiltonian~\cite{Rabi1937} extended to two resonators,
\bea
\hat H_{\rm R} &=& \omega_a \hat a^\dag \hat a + \omega_b \hat b^\dag \hat b + \frac{\omega_q}{2}\sz \nn\\
&&+ \left[g_a \left(\hat a + \hat a^\dag \right) + g_b \left(\hat b + \hat b^\dag \right) \right]\sx.
\label{eq:HRabi}
\eea

\begin{figure}
\centerline{
  \resizebox{\columnwidth}{!}{
    \begin{tikzpicture}[
      scale=0.55,
      level/.style={thick},
      virtual/.style={thick,densely dashed},
      ztrans/.style={thick,->,shorten >=0.2cm,shorten <=0.2cm,>=stealth,densely dashed,color=red},
      nrtrans/.style={thick,->,shorten >=0.2cm,shorten <=0.2cm,>=stealth,densely dashed,color=blue},
      rtrans/.style={thick,->,shorten >=0.2cm,shorten <=0.2cm,>=stealth,color=blue},
      classical/.style={thin,double,->,shorten >=4pt,shorten <=4pt,>=stealth}
    ]
    \draw[level] (2cm,0cm) -- (0cm,0cm) node[left] {$\ket{1,0,g}$};
    \draw[level] (2.5cm,3cm) -- (4.5cm,3cm) node[midway,above] {$\ket{1,1,e}$};
    \draw[level] (2.5cm,-4cm) -- (4.5cm,-4cm) node[midway,below] {$\ket{0,0,e}$};
    \draw[level] (5cm,4cm) -- (7cm,4cm) node[midway,above] {$\ket{1,2,g}$};
    \draw[level] (5cm,-3cm) -- (7cm,-3cm) node[midway,below] {$\ket{0,1,g}$};
    \draw[level] (7.5cm,0cm) -- (9.5cm,0cm) node[right] {$\ket{0,2,e}$};
    \draw[nrtrans] (1cm,0cm) -- (3.3cm,3cm) node[midway, left, xshift = -0.05cm, yshift = 0cm] {$g_b \hat b^\dag \sp$};
    \draw[rtrans] (1cm,0cm) -- (3.1cm,-4cm) node[midway, left, xshift = 0cm, yshift = 0cm] {$g_a \hat a \sp$};
    \draw[rtrans] (4.1cm,3cm) -- (5.5cm,4cm) node[midway, right, xshift = -0.05cm, yshift = -0.1cm] {$g_b \hat b^\dag \sm$};
    \draw[nrtrans] (4.1cm,3cm) -- (5.7cm,-3cm) node[midway, right, xshift = 0cm, yshift = 0cm] {$g_a \hat a \sm$};
    \draw[rtrans] (3.5cm,-4cm) -- (5.4cm,-3cm) node[above, left, xshift = -0.15cm, yshift = 0.05cm] {$g_b \hat b^\dag \sm$};
    \draw[rtrans] (6.5cm,4cm) -- (8.5cm,0cm) node[midway, right, xshift = 0cm, yshift = 0cm] {$g_a \hat a \sp$};
    \draw[nrtrans] (6.5cm,-3cm) -- (8.5cm,0cm) node[midway, right, xshift = 0cm, yshift = 0cm] {$g_b \hat b^\dag \sp$};
    \end{tikzpicture}
  }
}
\caption{The lowest-order processes contributing to a transition between $\ket{1,0,g}$ and $\ket{0,2,e}$ in the quantum Rabi model. The transitions that do not conserve excitation number are shown as dashed blue lines and the excitation-number-conserving transitions are shown as solid blue lines. The label of each line is the term in \eqref{eq:HRabi} that gives rise to that transition. The parameters $\omega_a = 5\omega_q$ and $\omega_b = 2\omega_q$ were used to set the positions of the energy levels. \label{fig:RabiProcesses10gTo02e}}
\end{figure}
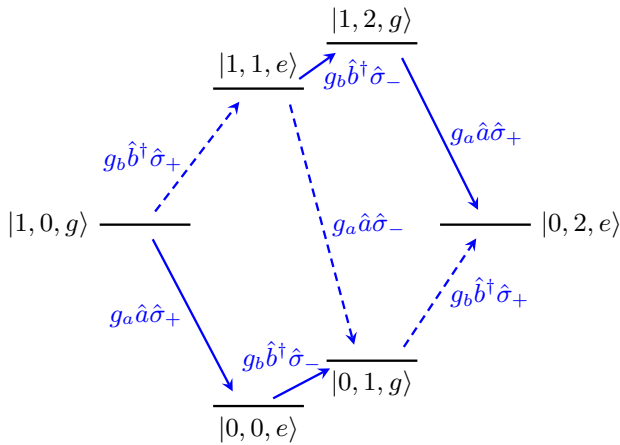

Placing the system close to the resonance $\omega_a = 2\omega_b + \omega_q$, virtual transitions involving the intermediate states $\ket{0,0,e}$, $\ket{0,1,g}$, $\ket{1,1,e}$, and $\ket{1,2,g}$ (to lowest order), contribute to the process $\ket{1,0,g} \leftrightarrow \ket{0,2,e}$, as shown in \figref{fig:RabiProcesses10gTo02e}. The most direct path between $\ket{1,0,g}$ and $\ket{0,2,e}$ involves three steps, since only one photon can be created or annihilated in each step. We note that all the paths include at least one transition that is due to terms in the Hamiltonian that do not conserve excitation number (dashed arrows in the figure).

Retaining only the states shown in \figref{fig:RabiProcesses10gTo02e}, we can write the quantum Rabi Hamiltonian from \eqref{eq:HRabi} on matrix form as
\begin{widetext}
\be
\hat H_{\rm R} = 
\begin{pmatrix}
\frac{\omega_q}{2} & g_b & g_a & 0 & 0 & 0 \\
g_b & \omega_b - \frac{\omega_q}{2} & 0 & \sqrt{2}g_b & g_a & 0 \\
g_a & 0 & \omega_a - \frac{\omega_q}{2} & 0 & g_b & 0 \\
0 & \sqrt{2}g_b & 0 & 2\omega_b + \frac{\omega_q}{2} & 0 & g_a \\
0 & g_a & g_b & 0 & \omega_a + \omega_b + \frac{\omega_q}{2} & \sqrt{2}g_b \\
0 & 0 & 0 & g_a & \sqrt{2}g_b & \omega_a + 2\omega_b - \frac{\omega_q}{2}
\end{pmatrix}
,
\label{eq:HTruncated10g02e}
\ee
\end{widetext}
where the states are ordered as $\ket{0,0,e}$, $\ket{0,1,g}$, $\ket{1,0,g}$, $\ket{0,2,e}$, $\ket{1,1,e}$, and $\ket{1,2,g}$. Just like in \secref{sec:SinglePhoton}, we can adiabatically eliminate the intermediate states when the condition $\omega_a \approx 2\omega_b + \omega_q$ is satisfied. The result of this calculation, the details of which are given in \appref{app:AnalyticalCalculations10g02e}, is an effective coupling between the states $\ket{1,0,g}$ and $\ket{0,2,e}$ with magnitude
\be
g_{\rm eff} = \frac{\sqrt{2} g^3 \left[ 2\omega_b (\omega_a - 2\omega_b) - g^2 \right]}{2\omega_b^2 (\omega_a - \omega_b)^2 + g^2 \omega_b (5\omega_b - 3\omega_a) + g^4}
\label{eq:Geff10g02eRabiEqualG}
\ee
on resonance. Here, we have set $g_a = g_b \equiv g$ to simplify the expression slightly. We note that, to leading order, the coupling scales like $g^3/\omega^2$; indeed, the leading-order term is
\be
g_{\rm eff} = \frac{\sqrt{2} g^3 \left(\omega_a - 2\omega_b \right)}{\omega_b \left(\omega_a - \omega_b \right)^2}.
\label{eq:Geff10g02eRabiEqualGLeadingOrder}
\ee
This is a factor $g/\omega$ weaker than for the single-photon frequency conversion, and reflects the fact that an additional intermediate transition is required for the two-photon conversion. We also note that the coupling becomes small in the limit of small $\omega_q$, i.e., when $2\omega_b \to \omega_a$. The coupling would become large if $\omega_a \to \omega_b$, but this is impossible since $\omega_a = 2\omega_b + \omega_q$ in this scheme.

\begin{figure*}
\centering
\includegraphics[width=\linewidth]{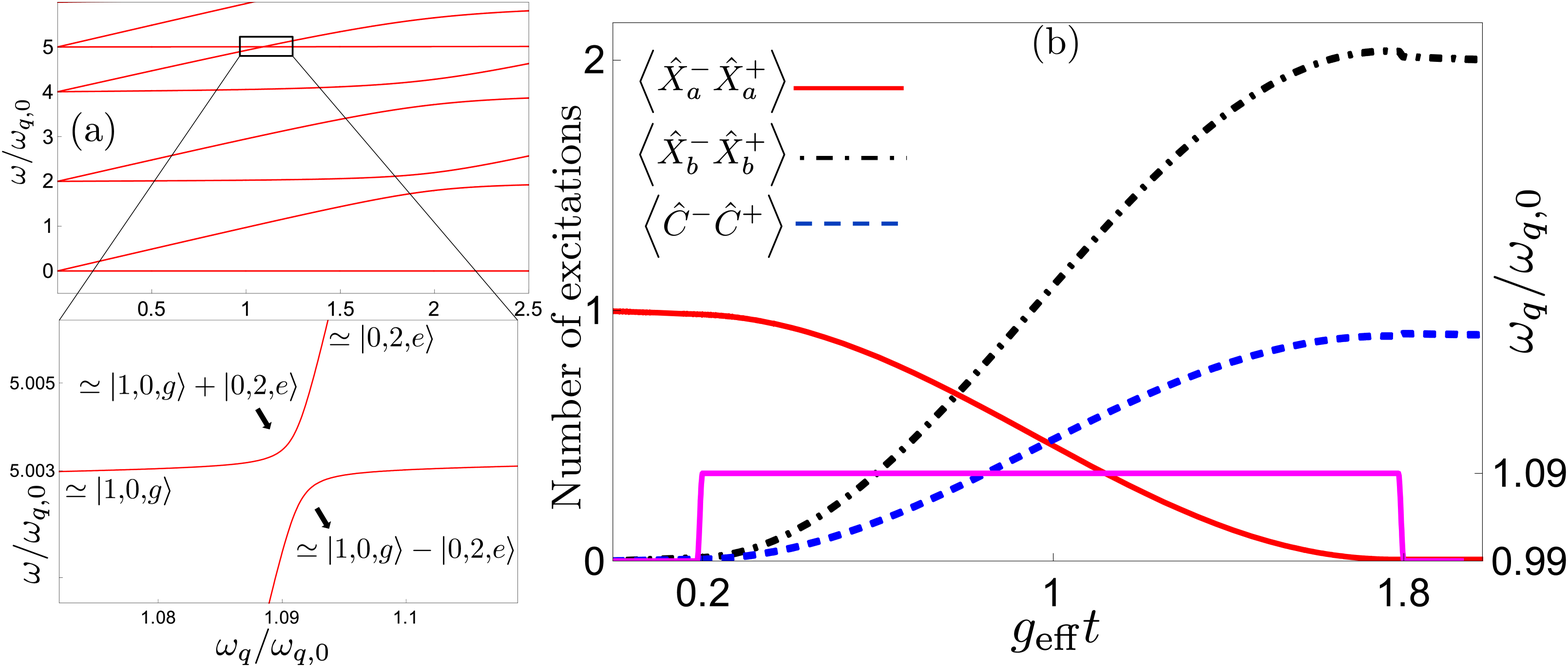}
\caption{Two-photon frequency conversion via transitions between $\ket{1,0,g}$ and $\ket{0,2,e}$. (a) The energy levels of our system, given in \eqref{eq:HRabi}, plotted as a function of the qubit frequency $\omega_q$, using the parameters $g_a = g_b = 0.2 \omega_{q,0}$, $\omega_a = 5\omega_{q,0}$, and $\omega_b = 2\omega_{q,0}$, where the reference point $\omega_{q,0}$ is set such that $\omega_a = 2\omega_b + \omega_{q,0}$. In the zoom-in, close to the resonance $\omega_a = 2\omega_b + \omega_q$, we see the anticrossing between $\ket{1,0,g}$ and $\ket{0,2,e}$ with the splitting $2g_{\rm eff}$ given by \eqref{eq:Geff10g02eRabiEqualG}. Up-conversion of a photon pair into a single photon, or down-conversion of a single photon into a photon pair, can be achieved by adiabatically tuning $\omega_q$ to follow one of the energy levels in the figure from $\ket{0,2,e}$ to $\ket{1,0,g}$, or vice versa. (b) A rapid frequency conversion can be achieved by starting in $\ket{1,0,g}$ or $\ket{0,2,e}$, far from the resonance $\omega_a = 2\omega_b + \omega_q$, tuning the qubit frequency (pink solid line) into resonance for half a Rabi period ($\pi/2g_{\rm eff}$) and then tuning it out of resonance again. The figure shows the number of excitations in the two resonators (red solid line for $a$, black dashed-dotted line for $b$) and the qubit (blue dashed line) during such a process, including decoherence in the form of relaxation from the resonators and the qubit. The parameters used for the decoherence are $\kappa_a = \kappa_b = \gamma = 2\times 10^{-5} \omega_{q,0}$. 
\label{fig:LevelsAndDynamicTransferSimSinglePhoton10g02e}}
\end{figure*}

\begin{figure}
\centering
\includegraphics[width=\columnwidth]{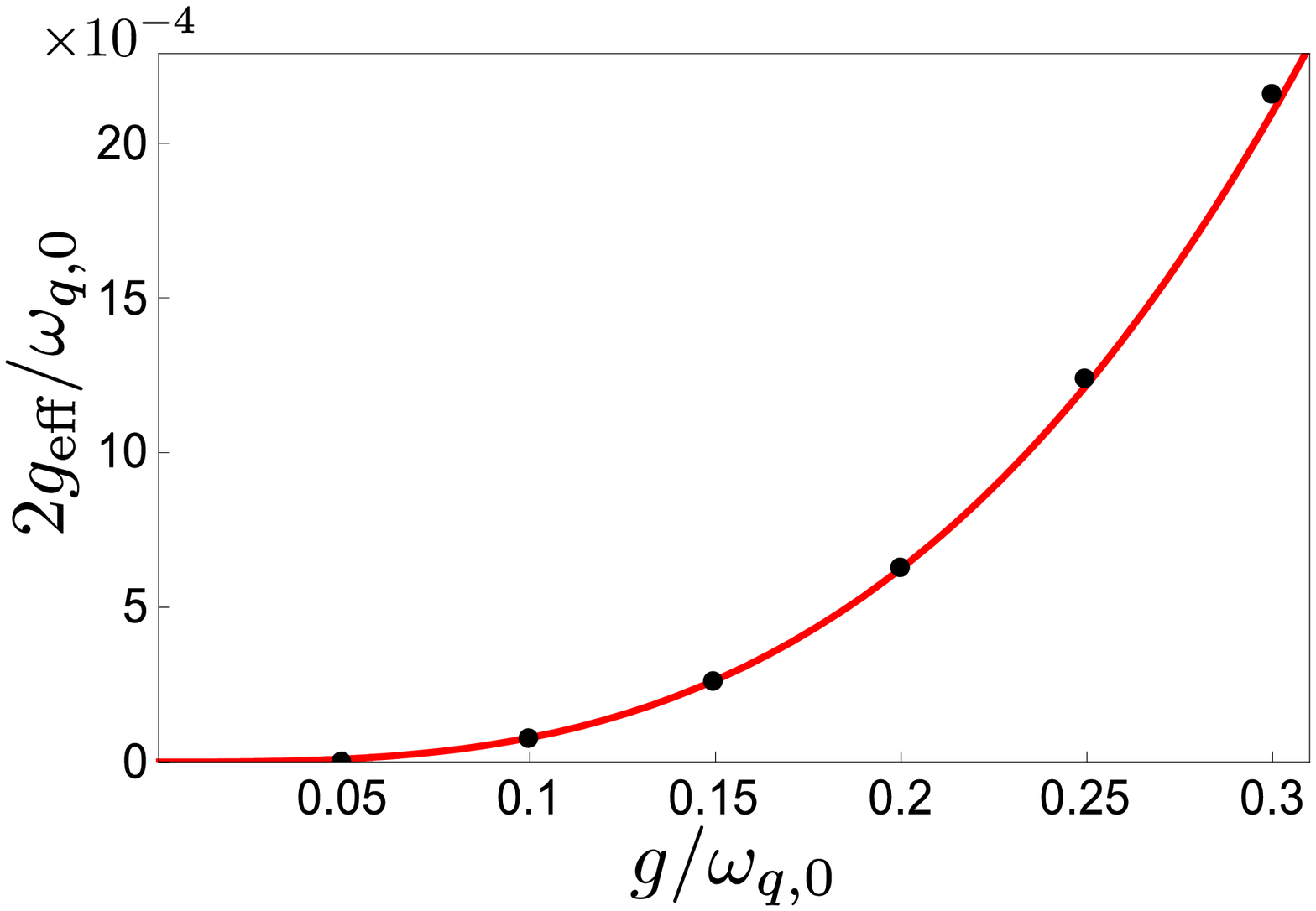}
\caption{Comparison of analytical (red line) and numerical (black dots) results for the effective coupling between the states $\ket{1,0,g}$ and $\ket{0,2,e}$. The graph shows the minimum energy splitting $2g_{\rm eff}/\omega_{q,0}$ as a function of $g/\omega_{q,0}$, using the same parameters as in \figref{fig:LevelsAndDynamicTransferSimSinglePhoton10g02e}. 
\label{fig:ComparingCoupling10g02e}}
\end{figure}

The two-photon frequency conversion can be performed either by adiabatic transfer or by tuning the qubit into resonance for half a Rabi oscillation period, as explained in \secref{sec:SinglePhoton}. In the first approach, one adiabatically tunes the qubit energy to follow one of the energy levels shown in \figref{fig:LevelsAndDynamicTransferSimSinglePhoton10g02e}(a). A simulation of the second approach, including decoherence, is shown in \figref{fig:LevelsAndDynamicTransferSimSinglePhoton10g02e}(b). The timescale for these processes is set by the effective coupling. In \figref{fig:ComparingCoupling10g02e}, we show that the expression for the effective coupling given in \eqref{eq:Geff10g02eRabiEqualG} remains a good approximation up to at least $g = 0.3\omega_{q,0}$ for the parameters used in \figref{fig:LevelsAndDynamicTransferSimSinglePhoton10g02e}.

\subsection{$\ket{1,0,e} \leftrightarrow \ket{0,2,g}$}

\begin{figure}
\centerline{
  \resizebox{\columnwidth}{!}{
    \begin{tikzpicture}[
      scale=0.55,
      level/.style={thick},
      virtual/.style={thick,densely dashed},
      ztrans/.style={thick,->,shorten >=0.2cm,shorten <=0.2cm,>=stealth,densely dashed,color=red},
      nrtrans/.style={thick,->,shorten >=0.2cm,shorten <=0.2cm,>=stealth,densely dashed,color=blue},
      rtrans/.style={thick,->,shorten >=0.2cm,shorten <=0.2cm,>=stealth,color=blue},
      classical/.style={thin,double,<->,shorten >=4pt,shorten <=4pt,>=stealth}
    ]
    \draw[level] (2cm,0cm) -- (0cm,0cm) node[left] {$\ket{1,0,e}$};
    \draw[level] (2.5cm,1cm) -- (4.5cm,1cm) node[midway,above] {$\ket{1,1,g}$};
    \draw[level] (2.5cm,-4cm) -- (4.5cm,-4cm) node[midway,below] {$\ket{0,0,g}$};
    \draw[level] (5cm,4cm) -- (7cm,4cm) node[midway,above] {$\ket{1,2,e}$};
    \draw[level] (5cm,-1cm) -- (7cm,-1cm) node[midway,below] {$\ket{0,1,e}$};
    \draw[level] (7.5cm,0cm) -- (9.5cm,0cm) node[right] {$\ket{0,2,g}$};
    \draw[rtrans] (1cm,0cm) -- (3.3cm,1cm) node[midway, left, xshift = 0.1cm, yshift = 0.2cm] {$g_b \hat b^\dag \sm$};
    \draw[nrtrans] (1cm,0cm) -- (3.3cm,-4cm) node[midway, left, xshift = 0cm, yshift = 0cm] {$g_a \hat a \sm$};
    \draw[nrtrans] (4.3cm,1cm) -- (6cm,4cm) node[midway, left, xshift = 0.05cm, yshift = 0.1cm] {$g_b \hat b^\dag \sp$};
    \draw[rtrans] (4.3cm,1cm) -- (5.8cm,-1cm) node[midway, left, xshift = 0.05cm, yshift = -0.05cm] {$g_a \hat a \sp$};
    \draw[nrtrans] (3.5cm,-4cm) -- (5.1cm,-1cm) node[midway, right, xshift = 0cm, yshift = 0cm] {$g_b \hat b^\dag \sp$};
    \draw[nrtrans] (6.3cm,4cm) -- (8.7cm,0cm) node[midway, right, xshift = 0.05cm, yshift = 0cm] {$g_a \hat a \sm$};
    \draw[rtrans] (6.3cm,-1cm) -- (8.4cm,0cm) node[midway, right, xshift = 0cm, yshift = -0.05cm] {$g_b \hat b^\dag \sm$};
    \end{tikzpicture}
  }
}
\caption{The lowest-order processes contributing to a transition between $\ket{1,0,e}$ and $\ket{0,2,g}$ in the quantum Rabi model. The transitions that do not conserve excitation number are shown as dashed blue lines and the excitation-number-conserving transitions are shown as solid blue lines. The label of each line is the term in \eqref{eq:HRabi} that gives rise to that transition. The parameters $\omega_a = 3\omega_q$ and $\omega_b = 2\omega_q$ were used to set the positions of the energy levels. \label{fig:RabiProcesses10eTo02g}}
\end{figure}
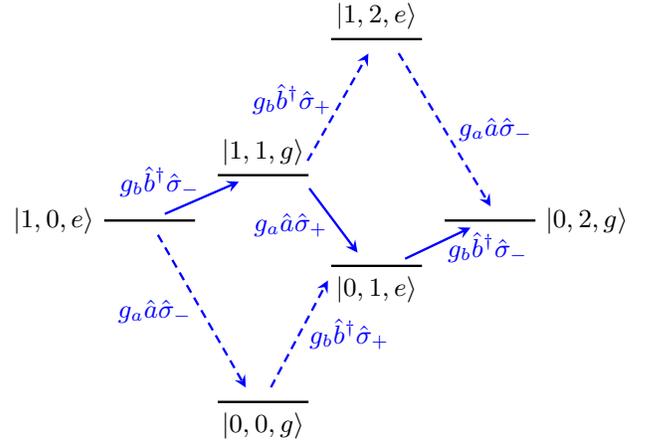

For the process $\ket{1,0,e} \leftrightarrow \ket{0,2,g}$, we show in \figref{fig:RabiProcesses10eTo02g} the virtual transitions from the quantum Rabi Hamiltonian that contribute to lowest order. We note that this process conserves the excitation number, which means that there is a path between the states that can be realized using only terms from the Jaynes--Cummings (JC) Hamiltonian~\cite{Jaynes1963} (solid arrows in the figure). Below, we analyze the effective coupling both for the full quantum Rabi Hamiltonian and for the JC Hamiltonian.

\subsubsection{Quantum Rabi Hamiltonian}
\label{sec:10e02gRabi}

Retaining only the states shown in \figref{fig:RabiProcesses10eTo02g}, we can write the quantum Rabi Hamiltonian from \eqref{eq:HRabi} on matrix form as
\begin{widetext}
\be
\hat H_{\rm R} = 
\begin{pmatrix}
- \frac{\omega_q}{2} & g_b & g_a & 0 & 0 & 0 \\
g_b & \omega_b + \frac{\omega_q}{2} & 0 & \sqrt{2}g_b & g_a & 0 \\
g_a & 0 & \omega_a + \frac{\omega_q}{2} & 0 & g_b & 0 \\
0 & \sqrt{2}g_b & 0 & 2\omega_b - \frac{\omega_q}{2} & 0 & g_a \\
0 & g_a & g_b & 0 & \omega_a + \omega_b - \frac{\omega_q}{2} & \sqrt{2}g_b \\
0 & 0 & 0 & g_a & \sqrt{2}g_b & \omega_a + 2\omega_b + \frac{\omega_q}{2}
\end{pmatrix}
,
\label{eq:HRabiTruncated10e02g}
\ee
\end{widetext}
where the states are ordered as $\ket{0,0,g}$, $\ket{0,1,e}$, $\ket{1,0,e}$, $\ket{0,2,g}$, $\ket{1,1,g}$, and $\ket{1,2,e}$. As in previous calculations, we can perform adiabatic elimination close to the resonance, which in this case is $\omega_a + \omega_q \approx 2\omega_b$. The details of the elimination are given in \appref{app:AnalyticalCalculations10e02gRabi}. The result is an effective coupling between the states $\ket{1,0,e}$ and $\ket{0,2,g}$ with magnitude
\be
g_{\rm eff} = \frac{\sqrt{2} g^3 \left[ 2\omega_b (\omega_a - 2\omega_b) - g^2 \right]}{2\omega_b^2 (\omega_a - \omega_b)^2 + g^2 \omega_b (5\omega_b - 3\omega_a) + g^4}
\label{eq:Geff10e02gRabiEqualG}
\ee
on resonance. We have set $g_a = g_b \equiv g$ to simplify the expression slightly. Note that this expression for the coupling is actually exactly the same as the one for the process $\ket{1,0,g} \leftrightarrow \ket{0,2,e}$ given in \eqref{eq:Geff10g02eRabiEqualG}. Even though the two processes use different intermediate states, the truncated Hamiltonians in Eqs.~(\ref{eq:HTruncated10g02e}) and (\ref{eq:HRabiTruncated10e02g}) only differ in the sign of $\omega_q$. Since $\omega_q$ is replaced on resonance by $(\omega_a - 2\omega_b)$ in the first case and by $(2\omega_b - \omega_a)$ in the second case, the formula for the effective coupling ends up being the same in both cases. The two cases still differ, however. For example, while the limit $\omega_a \to \omega_b$, which enhances the coupling, could not occur for the process $\ket{1,0,g} \leftrightarrow \ket{0,2,e}$, it is possible for $\ket{1,0,e} \leftrightarrow \ket{0,2,g}$. However, in this limit the approximations behind the adiabatic elimination break down, since the states $\ket{1,1,g}$ and $\ket{0,1,e}$ would also be on resonance and become populated.

\begin{figure*}
\centering
\includegraphics[width=\linewidth]{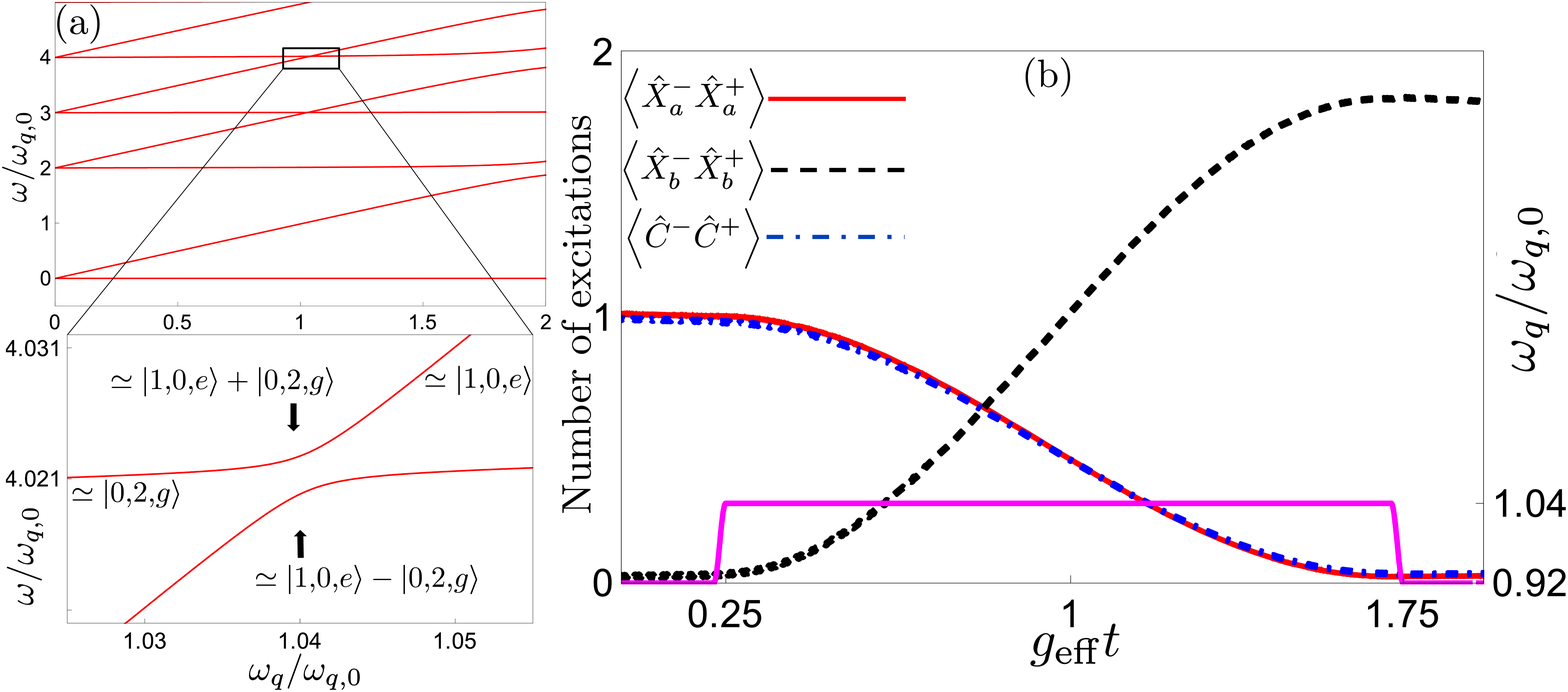}
\caption{Two-photon frequency conversion via transitions between $\ket{1,0,e}$ and $\ket{0,2,g}$. (a) The energy levels of our system, given in \eqref{eq:HRabi}, plotted as a function of the qubit frequency $\omega_q$, using the parameters $g_a = g_b = 0.125 \omega_{q,0}$, $\omega_a = 3\omega_{q,0}$, and $\omega_b = 2\omega_{q,0}$, where the reference point $\omega_{q,0}$ is set such that $\omega_a + \omega_{q,0} = 2\omega_b$. In the zoom-in, close to the resonance $\omega_a + \omega_q = 2\omega_b$, we see the anticrossing between $\ket{1,0,e}$ and $\ket{0,2,g}$ with the splitting $2g_{\rm eff}$ given by \eqref{eq:Geff10e02gRabiEqualG}. Up-conversion of a photon pair into a single photon, or down-conversion of a single photon into a photon pair, can be achieved by adiabatically tuning $\omega_q$ to follow one of the energy levels in the figure from $\ket{0,2,g}$ to $\ket{1,0,e}$, or vice versa. (b) A rapid frequency conversion can be achieved by starting in $\ket{1,0,e}$ or $\ket{0,2,g}$, far from the resonance $\omega_a + \omega_q = 2\omega_b$, tuning the qubit frequency (pink solid line) into resonance for half a Rabi period ($\pi/2g_{\rm eff}$) and then tuning it out of resonance again. The figure shows the number of excitations in the two resonators (red solid line for $a$, black dashed-dotted line for $b$) and the qubit (blue dashed line) during such a process, including decoherence in the form of relaxation from the resonators and the qubit. The parameters used for the decoherence are $\kappa_a = \kappa_b = \gamma = 4\times 10^{-5}\omega_{q,0}$. 
\label{fig:LevelsAndDynamicTransferSimSinglePhoton10e02g}}
\end{figure*}

The two-photon frequency conversion can again be performed either by adiabatic transfer or by tuning the qubit into resonance for half a Rabi oscillation period, as explained in \secref{sec:SinglePhoton}. The energy levels to follow in the first approach are plotted in \figref{fig:LevelsAndDynamicTransferSimSinglePhoton10e02g}(a) and a simulation of the second approach, including decoherence, is shown in \figref{fig:LevelsAndDynamicTransferSimSinglePhoton10e02g}(b).

\subsubsection{Jaynes--Cummings Hamiltonian}

For completeness, we calculate the effective coupling using only the JC Hamiltonian for two resonators and one qubit, i.e., we eliminate the non-excitation-conserving terms in the quantum Rabi Hamiltonian of \eqref{eq:HRabi} using the RWA, giving
\bea
\hat H_{\rm JC} &=& \omega_a \hat a^\dag \hat a + \omega_b \hat b^\dag \hat b + \frac{\omega_q}{2}\sz \nn\\
&&+ g_a \left(\hat a\sp + \hat a^\dag\sm \right) + g_b \left(\hat b\sp + \hat b^\dag\sm \right).
\label{eq:HJC}
\eea

Retaining only the states connected by solid arrows in \figref{fig:RabiProcesses10eTo02g}, we can write the Hamiltonian from \eqref{eq:HJC} on matrix form as
\bea
\hat H_{\rm JC} = 
\begin{pmatrix}
\omega_b + \frac{\omega_q}{2} & 0 & \sqrt{2}g_b & g_a \\
0 & \omega_a + \frac{\omega_q}{2} & 0 & g_b \\
\sqrt{2}g_b & 0 & 2\omega_b - \frac{\omega_q}{2} & 0 \\
g_a & g_b & 0 & \omega_a + \omega_b - \frac{\omega_q}{2}
\end{pmatrix},\nn\\
\label{eq:HJCTruncated}
\eea
where the states are ordered as $\ket{0,1,e}$, $\ket{1,0,e}$, $\ket{0,2,g}$, and $\ket{1,1,g}$. Again, we perform adiabatic elimination close to the resonance $\omega_a + \omega_q \approx 2\omega_b$. The details of the elimination are given in \appref{app:AnalyticalCalculations10e02gJC}. The result is an effective coupling between the states $\ket{1,0,e}$ and $\ket{0,2,g}$ with magnitude
\be
g_{\rm eff} = - \frac{\sqrt{2}g_a g_b^2}{g_a^2 + (\omega_a - \omega_b)^2}
\label{eq:Geff10eTo02gJC}
\ee
on resonance. Just as for the other two-photon frequency-conversion processes, the coupling scales like $g^3/\omega^2$ to leading order. In fact, \eqref{eq:Geff10eTo02gJC} is a good approximation to \eqref{eq:Geff10e02gRabiEqualG}, since the path given by the JC terms (solid lines) in \figref{fig:RabiProcesses10eTo02g} is far less detuned in energy from the initial and final states than all the other paths and thus gives the largest contribution to the result in \eqref{eq:Geff10e02gRabiEqualG}. The remarks on the limit $\omega_a \to \omega_b$ given in \secref{sec:10e02gRabi} apply here as well. The schemes for implementing the frequency conversion are already given in \figref{fig:LevelsAndDynamicTransferSimSinglePhoton10e02g}.

\begin{figure}
\centering
\includegraphics[width=\linewidth]{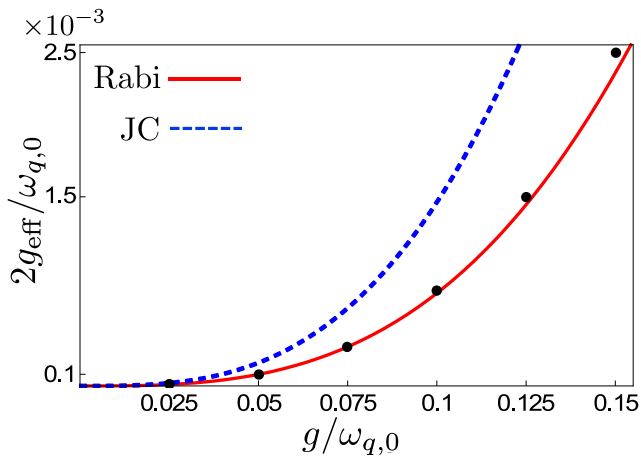}
\caption{Comparison of analytical (JC Hamiltonian [dashed blue line] and quantum Rabi Hamiltonian [solid red line]) and numerical (black dots) results for the effective coupling between the states $\ket{1,0,e}$ and $\ket{0,2,g}$. The graph shows the minimum energy splitting $2g_{\rm eff}/\omega_{q,0}$ as a function of $g/\omega_{q,0}$, using the same parameters as in \figref{fig:LevelsAndDynamicTransferSimSinglePhoton10e02g}. 
\label{fig:ComparingCoupling10e02g}}
\end{figure}

In \figref{fig:ComparingCoupling10e02g}, we compare the results from Eqs.~(\ref{eq:Geff10e02gRabiEqualG}) and (\ref{eq:Geff10eTo02gJC}) with a full numerical calculation. The contribution from the JC part dominates the coupling up until around $g_a = g_b = g = 0.03 \omega_{q,0}$ and gives a good approximation until then. For higher values of the coupling, using the approximation from the quantum Rabi Hamiltonian instead works fine up until around $g_a = g_b = g = 0.15\omega_{q,0}$.

\section{Summary and outlook}
\label{sec:SummaryOutlook}

We have shown how a system consisting of two resonators ultrastrongly coupled to a qubit can be used to realize a variety of frequency-conversion processes. In particular, we have shown how to convert a single photon into another photon of either higher or lower frequency, as well as how to convert a single photon into a photon pair and vice versa. All these processes are deterministic, can be implemented within a single setup, and do not require any external drives. The conversion is controlled by tuning the frequency of the qubit to and from values that make the desired transitions resonant.

Given the recent advances in USC circuit QED, we believe that our proposal can be implemented in such a setup. Indeed, two resonators have already been ultrastrongly coupled to a superconducting flux qubit~\cite{Baust2016}. Also, our proposal does not require very high coupling strengths. We only need that $g^2/\omega$ is appreciable (larger than the relevant decoherence rates) to realize single-photon frequency conversion; multi-photon frequency conversion can be demonstrated if $g^3/\omega^2$ is large enough.

A straightforward extension of the current work is to extend the calculations to processes with more photons in the second resonator or to add more resonators to the setup. Some of these possibilities are discussed in Ref.~\cite{Kockum2017a}, where we explore analogies of nonlinear optics in USC systems, including the fact that the processes in the current work can be considered analogies of Raman and hyper-Raman scattering if the qubit is thought of as playing the role of a phonon. More general three-wave mixing, such as $\ket{1,0,0,e} \leftrightarrow \ket{0,1,1,g}$, or third-harmonic and -subharmonic generation such as $\ket{1,0,e} \leftrightarrow \ket{0,3,g}$, are examples of schemes that can be considered, but it must be kept in mind that higher-order processes with more photons involved will have lower effective coupling strengths. Another direction for future work is to investigate how the precise qubit control of the frequency-conversion processes discussed here can be used to prepare photon bundles~\cite{SanchezMunoz2014} or interesting quantum superposition states with photons of different frequencies, a topic currently being explored in several frequency ranges~\cite{Chirolli2010, Zakka-Bajjani2011, Clemmen2016}.

\section*{Acknowledgements}

We acknowledge useful discussions with Roberto Stassi and Adam Miranowicz. This work is partially supported by the RIKEN iTHES Project, the MURI Center for Dynamic Magneto-Optics via the AFOSR award number FA9550-14-1-0040, the IMPACT program of JST, CREST, a Grant-in-Aid for Scientific Research (A), and from the MPNS COST Action MP1403 Nanoscale Quantum Optics. A.F.K. acknowledges support from a JSPS Postdoctoral Fellowship for Overseas Researchers.

\newpage
\appendix

\section{Analytical calculations of conversion rates}
\label{app:AnalyticalCalculations}

In this appendix, we present the full adiabatic-elimination calculations for the effective couplings in the three processes considered in this article: $\ket{1,0,g} \leftrightarrow \ket{0,1,e}$, $\ket{1,0,g} \leftrightarrow \ket{0,2,e}$, and $\ket{1,0,e} \leftrightarrow \ket{0,2,g}$. For each case, we compare the analytical results with numerical simulations to determine in what parameter regimes the analytical calculations constitute a good approximation.

\subsection{$\ket{1,0,g} \leftrightarrow \ket{0,1,e}$}
\label{app:AnalyticalCalculationsSinglePhoton}

Starting from the truncated Hamiltonian in \eqref{eq:HTruncatedSinglePhoton}, we move to a frame rotating with $(\omega_a - \frac{\omega_q}{2})$, i.e., subtracting $(\omega_a - \frac{\omega_q}{2})$ from the diagonal of the Hamiltonian, giving
\begin{widetext}
\bea
\hat H = 
\begin{pmatrix}
- \omega_a & 0 & - g_a\sin\theta & g_b\cos\theta & 0 & 0 \\
0 & -(\omega_a - \omega_q) & g_a\cos\theta & g_b\sin\theta & 0 & 0 \\
- g_a\sin\theta & g_a\cos\theta & 0 & 0 & - g_b\sin\theta & g_b\cos\theta \\
g_b\cos\theta & g_b\sin\theta & 0 & \omega_b + \omega_q - \omega_a & g_a\cos\theta & g_a\sin\theta \\
0 & 0 & - g_b\sin\theta & g_a\cos\theta & \omega_b & 0 \\
0 & 0 & g_b\cos\theta & g_a\sin\theta & 0 & \omega_b + \omega_q
\end{pmatrix}.
\eea
\end{widetext}
Denoting the amplitudes of the six states by $c_1$--$c_6$, respectively, the Schr\"odinger equation gives
\bea
i \dot c_1 &=& -\omega_a c_1 - g_a\sin\theta c_3 + g_b\cos\theta c_4, \\
i \dot c_2 &=& -(\omega_a - \omega_q) c_2 + g_a\cos\theta c_3 + g_b\sin\theta c_4, \\
i \dot c_3 &=& - g_a\sin\theta c_1 + g_a\cos\theta c_2 - g_b\sin\theta c_5 \nn\\
&&+ g_b\cos\theta c_6, \\
i \dot c_4 &=& (\omega_b + \omega_q - \omega_a) c_4 + g_b\cos\theta c_1 + g_b\sin\theta c_2 \nn\\
&&+ g_a\cos\theta c_5 + g_a\sin\theta c_6, \\
i \dot c_5 &=& \omega_b c_5 - g_b\sin\theta c_3 + g_a\cos\theta c_4, \\
i \dot c_6 &=& (\omega_b + \omega_q) c_6 + g_b\cos\theta c_3 + g_a\sin\theta c_4.
\eea
Assuming that $\omega_a \approx \omega_b + \omega_q$, and that $g_a, g_b \ll \omega_a, \omega_b, \abs{\omega_a - \omega_q}, \omega_b + \omega_q$, we can adiabatically eliminate the four intermediate levels (their population will not change significantly), i.e., set $\dot c_1 = \dot c_2 = \dot c_5 = \dot c_6 = 0$. This gives
\bea
c_1 &=& -\frac{g_a\sin\theta}{\omega_a} c_3 + \frac{g_b\cos\theta}{\omega_a} c_4, \\
c_2 &=& \frac{g_a\cos\theta}{\omega_a - \omega_q} c_3 + \frac{g_b\sin\theta}{\omega_a - \omega_q} c_4, \\
c_5 &=& \frac{g_b\sin\theta}{\omega_b} c_3 - \frac{g_a\cos\theta}{\omega_b} c_4, \\
c_6 &=& - \frac{g_b\cos\theta}{\omega_b + \omega_q} c_3 - \frac{g_a\sin\theta}{\omega_b + \omega_q} c_4,
\eea
which we then insert into the equations for $c_3$ and $c_4$ to arrive at
\bea
i \dot c_3 &=& \left( \frac{g_a^2 \sin^2\theta}{\omega_a} + \frac{g_a^2 \cos^2\theta}{\omega_a - \omega_q} - \frac{g_b^2 \sin^2 \theta}{\omega_b} - \frac{g_b^2 \cos^2 \theta}{\omega_b + \omega_q} \right) c_3 \nn\\
&&+ \frac{1}{2} g_a g_b\sin2\theta \nn\\
&&\times \left( \frac{1}{\omega_a-\omega_q} + \frac{1}{\omega_b} - \frac{1}{\omega_a} - \frac{1}{\omega_b+\omega_q}\right) c_4,
\eea
\bea
i \dot c_4 &=& \frac{1}{2} g_a g_b\sin2\theta \left( \frac{1}{\omega_a-\omega_q} + \frac{1}{\omega_b} - \frac{1}{\omega_a} - \frac{1}{\omega_b+\omega_q}\right) c_3 \nn\\
&&+ \bigg( \omega_b + \omega_q - \omega_a  + \frac{g_b^2 \cos^2\theta}{\omega_a} + \frac{g_b^2\sin^2\theta}{\omega_a - \omega_q} \nn\\
&& - \frac{g_a^2\cos^2 \theta}{\omega_b} - \frac{g_a\sin\theta}{\omega_b + \omega_q} \bigg) c_4.
\eea
While the energy level shifts in these equations are not final (they can be affected by processes involving more energy levels), the effective coupling rate between $\ket{1,0,g}$ and $\ket{0,1,e}$ is shown to be
\be
g_{\rm eff} = \frac{1}{2} g_a g_b\sin2\theta \left( \frac{1}{\omega_a-\omega_q} + \frac{1}{\omega_b} - \frac{1}{\omega_a} - \frac{1}{\omega_b+\omega_q}\right).
\label{eq:GeffSinglePhotonNotSimplified}
\ee
Assuming that we are exactly on resonance, the qubit frequency can be eliminated from this expression using $\omega_q = \omega_a - \omega_b$, leading to
\be
g_{\rm eff} = g_a g_b\sin2\theta \left( \frac{1}{\omega_b} - \frac{1}{\omega_a}\right) = \frac{g_a g_b(\omega_a - \omega_b)\sin2\theta}{\omega_a\omega_b},
\ee
the first part of which is given in \eqref{eq:GeffSinglePhoton}. We note that the result agrees with the perturbation-theory calculations performed in Ref.~\cite{Kockum2017a}. In general, the adiabatic elimination is more exact, but for a second-order process the result for the effective coupling is the same with both methods.

\subsection{$\ket{1,0,g} \leftrightarrow \ket{0,2,e}$}
\label{app:AnalyticalCalculations10g02e}

Starting from the truncated Hamiltonian in \eqref{eq:HTruncated10g02e}, we move to a frame rotating with $(\omega_a - \frac{\omega_q}{2})$, i.e., subtracting $(\omega_a - \frac{\omega_q}{2})$ from the diagonal of the Hamiltonian, giving
\begin{widetext}
\bea
\hat H_{\rm R} =
\begin{pmatrix}
\omega_q - \omega_a & g_b & g_a & 0 & 0 & 0 \\
g_b & \omega_b - \omega_a & 0 & \sqrt{2}g_b & g_a & 0 \\
g_a & 0 & 0 & 0 & g_b & 0 \\
0 & \sqrt{2}g_b & 0 & 2\omega_b - \omega_a + \omega_q & 0 & g_a \\
0 & g_a & g_b & 0 & \omega_b + \omega_q & \sqrt{2}g_b \\
0 & 0 & 0 & g_a & \sqrt{2}g_b & 2\omega_b
\end{pmatrix}.
\label{eq:HmatrixRot10gTo02eRabi}
\eea
Denoting the amplitudes of the six states by $c_1$--$c_6$, the Schr\"odinger equation gives
\bea
i \dot c_1 &=& (\omega_q - \omega_a) c_1 + g_b c_2 + g_a c_3, \\
i \dot c_2 &=& (\omega_b - \omega_a) c_2 + g_b c_1 + \sqrt{2}g_b c_4 + g_a c_5, \\
i \dot c_3 &=& g_a c_1 + g_b c_5, \\
i \dot c_4 &=& (2\omega_b - \omega_a + \omega_q) c_4 + \sqrt{2}g_b c_2 + g_a c_6, \\
i \dot c_5 &=& (\omega_b + \omega_q) c_5 + g_a c_2 + g_b c_3 +  \sqrt{2}g_b c_6, \\
i \dot c_6 &=& 2\omega_b c_6 + g_a c_4 +  \sqrt{2}g_b c_5.
\eea
Assuming that $\omega_a \approx 2 \omega_b + \omega_q$, and that $g_a, g_b \ll \omega_b + \omega_q, \abs{\omega_b - \omega_q}, \abs{\omega_a - \omega_q}$, we can adiabatically eliminate the four intermediate levels, i.e., set $\dot c_1 = \dot c_2 = \dot c_5 = \dot c_6 = 0$. This gives
\bea
i \dot c_3 &=& \frac{g_a^4 \omega_b + g_a^2 \left[ \omega_b (\omega_a - \omega_b)^2 - g_b^2 (\omega_a + \omega_b) \right] + g_b^2 \omega_b \left[ g_b^2 + 2\omega_b (\omega_b - \omega_a) \right]}{2\omega_b^2 \left[ g_a^2 + (\omega_a - \omega_b)^2 \right] + g_b^4 + 3g_b^2 \omega_b (\omega_b - \omega_a)} c_3 \nn\\
&&+ \frac{g_a g_b^2 \left[ g_a^2 - 3g_b^2 + 4\omega_b (\omega_a - 2\omega_b) \right]}{\sqrt{2}\left\{ 2\omega_b^2 \left[ g_a^2 + (\omega_a - \omega_b)^2 \right] + g_b^4 + 3g_b^2 \omega_b (\omega_b - \omega_a) \right\}} c_4,
\eea
\end{widetext}
where we simplified the expressions somewhat by setting $\omega_q = \omega_a - 2 \omega_b$. While the energy level shift in this equation is not final (they can be affected by processes involving more energy levels), the effective coupling rate between $\ket{1,0,g}$ and $\ket{0,2,e}$ is shown to be
\be
g_{\rm eff} = \frac{g_a g_b^2 \left[ g_a^2 - 3g_b^2 + 4\omega_b (\omega_a - 2\omega_b) \right]}{\sqrt{2}\left\{ 2\omega_b^2 \left[ g_a^2 + (\omega_a - \omega_b)^2 \right] + g_b^4 + 3g_b^2 \omega_b (\omega_b - \omega_a) \right\}}.
\label{eq:Geff10g02eRabi}
\ee
Setting $g_a = g_b \equiv g$, this reduces to
\be
g_{\rm eff} = \frac{\sqrt{2} g^3 \left[ 2\omega_b (\omega_a - 2\omega_b) - g^2 \right]}{2\omega_b^2 (\omega_a - \omega_b)^2 + g^2 \omega_b (5\omega_b - 3\omega_a) + g^4},
\ee
which is \eqref{eq:Geff10g02eRabiEqualG}. As shown in \figref{fig:ComparingCoupling10g02e}, this expression for the effective coupling is a good approximation to the exact value up to at least $g_a = g_b = g = 0.3 \omega_{q,0}$. We can simplify the expression for the coupling further by only keeping terms to leading order in $g/\omega$; the result is
\be
g_{\rm eff} = \frac{\sqrt{2} g^3 \left(\omega_a - 2\omega_b \right)}{\omega_b \left(\omega_a - \omega_b \right)^2},
\ee
which is \eqref{eq:Geff10g02eRabiEqualGLeadingOrder}. This agrees with the perturbation-theory calculation in Ref.~\cite{Kockum2017a}, which only captures the leading-order term.

\subsection{$\ket{1,0,e} \leftrightarrow \ket{0,2,g}$}
\label{app:AnalyticalCalculations10e02g}

For the process $\ket{1,0,e} \leftrightarrow \ket{0,2,g}$, we perform adiabatic elimination starting from both the quantum Rabi Hamiltonian and the JC Hamiltonian.

\subsubsection{Quantum Rabi Hamiltonian}
\label{app:AnalyticalCalculations10e02gRabi}

Starting from the truncated Hamiltonian in \eqref{eq:HRabiTruncated10e02g}, we move to a frame rotating with $(\omega_a + \frac{\omega_q}{2})$, i.e., subtracting $(\omega_a + \frac{\omega_q}{2})$ from the diagonal of the Hamiltonian, giving
\begin{widetext}
\bea
\hat H_{\rm R} =
\begin{pmatrix}
- \omega_a - \omega_q & g_b & g_a & 0 & 0 & 0 \\
g_b & \omega_b - \omega_a & 0 & \sqrt{2}g_b & g_a & 0 \\
g_a & 0 & 0 & 0 & g_b & 0 \\
0 & \sqrt{2}g_b & 0 & 2\omega_b - \omega_a - \omega_q & 0 & g_a \\
0 & g_a & g_b & 0 & \omega_b - \omega_q & \sqrt{2}g_b \\
0 & 0 & 0 & g_a & \sqrt{2}g_b & 2\omega_b
\end{pmatrix}.
\label{eq:HmatrixRot10eTo02gRabi}
\eea
Denoting the amplitudes of the six states by $c_1$--$c_6$, the Schr\"odinger equation gives
\bea
i \dot c_1 &=& -(\omega_a + \omega_q) c_1 + g_b c_2 + g_a c_3, \\
i \dot c_2 &=& (\omega_b - \omega_a) c_2 + g_b c_1 + \sqrt{2}g_b c_4 + g_a c_5, \\
i \dot c_3 &=& g_a c_1 + g_b c_5, \\
i \dot c_4 &=& (2\omega_b - \omega_a - \omega_q) c_4 +  \sqrt{2}g_b c_2 + g_a c_6, \\
i \dot c_5 &=& (\omega_b - \omega_q) c_5 + g_a c_2 + g_b c_3 +  \sqrt{2}g_b c_6, \\
i \dot c_6 &=& 2\omega_b c_6 + g_a c_4 +  \sqrt{2}g_b c_5.
\eea
Assuming that $\omega_a + \omega_q \approx 2 \omega_b$, and that $g_a, g_b \ll \abs{\omega_b - \omega_q}, \omega_b + \omega_q, \omega_a + \omega_q, \abs{\omega_a - \omega_q}$, we can adiabatically eliminate the four intermediate levels, i.e., set $\dot c_1 = \dot c_2 = \dot c_5 = \dot c_6 = 0$. This gives
\bea
i \dot c_3 &=& \frac{g_a^4 \omega_b + g_a^2 \left[ \omega_b (\omega_a - \omega_b)^2 - g_b^2 (\omega_a + \omega_b) \right] + g_b^2 \omega_b \left[ g_b^2 + 2\omega_b (\omega_b - \omega_a) \right]}{2\omega_b^2 \left[ g_a^2 + (\omega_a - \omega_b)^2 \right] + g_b^4 + 3g_b^2 \omega_b (\omega_b - \omega_a)} c_3 \nn\\
&&+ \frac{g_a g_b^2 \left[ g_a^2 - 3g_b^2 + 4\omega_b (\omega_a - 2\omega_b) \right]}{\sqrt{2}\left\{ 2\omega_b^2 \left[ g_a^2 + (\omega_a - \omega_b)^2 \right] + g_b^4 + 3g_b^2 \omega_b (\omega_b - \omega_a) \right\}} c_4,
\eea
\end{widetext}
where we simplified the expressions somewhat by setting $\omega_q = 2\omega_b - \omega_a$. While the energy level shift in this equation is not final (it can be affected by processes involving more energy levels), the effective coupling rate between $\ket{1,0,e}$ and $\ket{0,2,g}$ is shown to be
\be
g_{\rm eff} = \frac{g_a g_b^2 \left[ g_a^2 - 3g_b^2 + 4\omega_b (\omega_a - 2\omega_b) \right]}{\sqrt{2}\left\{ 2\omega_b^2 \left[ g_a^2 + (\omega_a - \omega_b)^2 \right] + g_b^4 + 3g_b^2 \omega_b (\omega_b - \omega_a) \right\}}.
\label{eq:Geff10eTo02gRabi}
\ee
Setting $g_a = g_b \equiv g$, this reduces to
\be
g_{\rm eff} = \frac{\sqrt{2} g^3 \left[ 2\omega_b (\omega_a - 2\omega_b) - g^2 \right]}{2\omega_b^2 (\omega_a - \omega_b)^2 + g^2 \omega_b (5\omega_b - 3\omega_a) + g^4},
\label{eq:Geff10eTo02gRabiEqualGAppendix}
\ee
which is \eqref{eq:Geff10e02gRabiEqualG}. As noted in the main text, this is equal to the coupling for the case $\ket{1,0,g} \leftrightarrow \ket{0,2,e}$, but other values of $\omega_a$ and $\omega_b$ are permitted in this case. In particular, the coupling can be increased by letting $\omega_a \to \omega_b$, but the approximations we have used here break down when $\abs{\omega_a - \omega_b}$ becomes comparable to $g$. Again, the result agrees with the perturbation-theory calculation in Ref.~\cite{Kockum2017a}, which only captures the leading-order term.

\subsubsection{Jaynes--Cummings Hamiltonian}
\label{app:AnalyticalCalculations10e02gJC}

Starting from the truncated Hamiltonian in \eqref{eq:HJCTruncated}, we move to a frame rotating with $(\omega_a + \frac{\omega_q}{2})$, i.e., subtracting $(\omega_a + \frac{\omega_q}{2})$ from the diagonal of the Hamiltonian, giving
\bea
\hat H_{\rm JC} = 
\begin{pmatrix}
\omega_b - \omega_a & 0 & \sqrt{2}g_b & g_a \\
0 & 0 & 0 & g_b \\
\sqrt{2}g_b & 0 & 2\omega_b - \omega_a - \omega_q & 0 \\
g_a & g_b & 0 & \omega_b - \omega_q
\end{pmatrix}. \nn\\
\eea
Denoting the amplitudes of the four states by $c_1$--$c_4$, the Schr\"odinger equation gives
\bea
i \dot c_1 &=& (\omega_b - \omega_a) c_1 + \sqrt{2}g_b c_3 + g_a c_4, \\
i \dot c_2 &=& g_b c_4, \\
i \dot c_3 &=& (2\omega_b - \omega_a - \omega_q) c_3 +  \sqrt{2}g_b c_1, \\
i \dot c_4 &=& (\omega_b - \omega_q) c_4 + g_a c_1 + g_b c_2.
\eea
Assuming that $\omega_a + \omega_q \approx 2 \omega_b$, and that $g_a, g_b \ll \omega_a, \omega_b, \omega_q$, we can adiabatically eliminate the two intermediate levels, i.e., set $\dot c_1 = \dot c_4 = 0$. This gives
\bea
i \dot c_2 &=& - \frac{g_b^2 (\omega_a - \omega_b)}{g_a^2 + (\omega_a - \omega_b)^2} c_2 - \frac{\sqrt{2}g_a g_b^2}{g_a^2 + (\omega_a - \omega_b)^2} c_3, \quad\quad
\eea
where we set $\omega_q = 2\omega_b - \omega_a$. While the energy level shift in this equation is not final (it can be affected by processes involving more energy levels), the effective coupling rate between $\ket{1,0,e}$ and $\ket{0,2,g}$ is shown to be
\be
g_{\rm eff} = - \frac{\sqrt{2}g_a g_b^2}{g_a^2 + (\omega_a - \omega_b)^2},
\ee
which is \eqref{eq:Geff10eTo02gJC}. Setting $g_a = g_b \equiv g$, this reduces to
\be
g_{\rm eff} = - \frac{\sqrt{2}g^3}{g^2 + (\omega_a - \omega_b)^2},
\label{eq:Geff10eTo02gJCEqualG}
\ee
which to leading order in $g/\omega$ becomes
\be
g_{\rm eff} = - \frac{\sqrt{2} g^3}{\left(\omega_a - \omega_b \right)^2},
\ee
agreeing with the perturbation-theory calculation of Ref.~\cite{Kockum2017a}.

\bibliography{FrequencyConversionRefs}

\end{document}